\documentclass[smallabstract,smallcaptions]{dccpaper}

\usepackage{epsfig}
\usepackage{citesort}
\usepackage{amsmath}
\usepackage{amssymb}
\usepackage{color}
\usepackage{url}

\newlength{\figurewidth}
\newlength{\smallfigurewidth}

\usepackage[utf8x]{inputenc}
\usepackage{lipsum}
\usepackage{xspace}

\newcommand{\wt}{wt\xspace}
\newcommand{\wts}{wts\xspace}

\newcommand{\baseline}{baseline\xspace}
\newcommand{\wtrle}{wtrle\xspace}
\newcommand{\wtmap}{wtmap\xspace}

\begin{document}

\title{{\bf \Large Compact Representations of Event Sequences}
\footnote{
Funded in part by European Union’s Horizon 2020 research and innovation
programme under the Marie Sklodowska-Curie grant agreement No 690941.
NB, GdB and TR supported by: CDTI, MINECO grant ITC-20151247; MINECO grants
TIN2016-78011-C4-1-R, TIN2016-77158-C4-3-R and TIN2013-46238-C4-3-R; and Xunta de
Galicia grant ED431C 2017/58 GRC.
DS supported by: MINECO (PGE and FEDER) grants TIN2016-77158-C4-3-R and
TIN2013-46801-C4-3-R, and Fondecyt-Conicyt grant number 1170497.
GN supported by Basal Funds FB0001, Conicyt, Chile.
}
}

\author{\small %
Nieves R. Brisaboa$^\ast$, Guillermo de Bernardo$^\ast$, Gonzalo
Navarro$^\dag$,  Tirso V. Rodeiro$^\ast$ \\  and Diego Seco$^\ddag$\\[0.5em]
{\small\begin{minipage}{\linewidth}\begin{center}
\begin{tabular}{ccc}
$^{\ast}$University of A Coru\~na & $^{\dag}$CeBiB and DCC
& $^{\ddag}$University of Concepci\'on \\
A Coru\~na, Spain & University of Chile & Concepci\'on, Chile \\
\url{brisaboa@udc.es} & Santiago, Chile &
\url{dseco@udec.cl}\\
\url{gdebernardo@udc.es} & \url{gnavarro@dcc.uchile.cl} & \\
\url{tirso.varela.rodeiro@udc.es} &&\\
\end{tabular}
\end{center}\end{minipage}}
}

\maketitle

\begin{abstract}
We introduce a new technique for the efficient management of large sequences of
multi-dimensional data, which takes advantage
of regularities that arise in real-world datasets and supports different
types of aggregation queries. More importantly, our representation is flexible in the sense that the relevant dimensions and queries may be used
to guide the construction process, easily providing a space-time tradeoff
depending on the relevant queries in the domain. We provide two alternative
representations for sequences of multidimensional data and describe
the techniques to efficiently store the datasets and to perform aggregation
queries over the compressed representation. We perform experimental evaluation
on realistic datasets, showing the space efficiency and query capabilities of our proposal.
\end{abstract}

\Section{Introduction}

In recent years, the amount of information produced in different domains has increased exponentially. In addition to transactional data in operational databases, other areas such as Geographic Information Systems (GIS) or bioinformatics produce huge amounts of information obtained from sensor networks or other tools. In many cases, the information produced is multi-dimensional, in the sense that each entry contains a set of features that can be regarded as dimensions. For instance, in a sales database, each transaction involves a salesman, a client, a product, etc. 

In many domains, this data collection is then used to answer different kinds of analytic queries. Frequently queries involve grouping and filtering operations on the data, by specific dimensions, to extract relevant information from cumulative values. This is performed for instance in OLAP systems~\cite{OLAP}, designed to provide information about aggregated values according to multiple characteristics or dimensions of the data. For instance, a system storing the activity log  of employees will
contain entries, including time, employee and activity. From this dataset, we can obtain the time devoted to activity X for one/all the employees in a period. While this information can be computed from the original data, in most cases the information is processed and stored in a separate data warehouse to answer those aggregation queries.
 
However, many times data has some kind of ordering that can be exploited. For example, a traditional data warehouse cannot answer how many times activity X (\textit{meeting} with a customer) has been followed by the activity Y (\textit{sell} a product). In a GIS dataset, we may have the elevations of roads. A traditional data warehouse could answer queries about how many Km. a road X traverses at elevation 1000m, 1050m or 1010m, but we could also be interested in knowing if those Kms. at elevation 1000m are contiguous (the road has no slopes) or if they are short sectors because the road has many changes of elevation. In these cases where the order of the values of some variable is relevant, typical data warehouses are not capable of answering all the relevant queries because they cannot exploit the sequence of events (or values of the variable of interest: activity or elevation in our examples).
 


In this paper we introduce a new representation for sequences of multi-dimensional data that is designed to efficiently solve aggregation queries, while keeping all the information of the original sequence. Our proposal is specifically designed to take advantage of repetitiveness in the original sequence. This repetitiveness occurs in many real-world sequences due to the \emph{locality} of the data. For instance, in the example of activities performed by employees, an employee is likely to perform the same activity in consecutive time instants. Our proposal provides a space/time tradeoff, allowing us to use only the space necessary to provide indexing capabilities in the relevant dimensions. Additionally, in many cases our representation can also be easily ``extended'' to add query capabilities in new dimensions without altering the original structure and features. Our proposal is based on the wavelet tree~\cite{GGV_WT}. We combine wavelet trees with other compact data structures in order to exploit regularities in the datasets and to perform different types of counting/aggregation queries. We experimentally evaluate our representations, comparing them with a more traditional alternative based on storing separately the original data sequence and building an indexed representation to answer aggregation queries.

\Section{Background}

Due to space constraints, we introduce here the compact data structures we use as building blocks of our solutions, but we refer to~\cite{Navarrobook} for details about them. 
An ubiquitous building block of most compact data structures is a bitmap with support for rank/select operations. Given a binary sequence of length $n$, a bitmap $B[1,n]$ provides access to any position $i$ of the sequence, denoted as $access(B,i)$, counts the number of occurrences of bit $v$ up to position $i$, $rank_v(B,i)$, and retrieves the position of the $j$-th occurrence of bit $v$, $select_v(B,j)$. All these operations can be supported in $O(1)$ time with $n+o(n)$ space~\cite{Jac89}. When the binary sequence is very sparse (i.e. the number of $1$s, $m$, is much smaller than $n$), space can be improved to $nH_0(B)+O(m)$ bits, supporting select in $O(1)$ time and rank in $O(min(\log m,\log\frac{n}{m}))$ time~\cite{OkanoharaS07}.

The \textit{wavelet tree}~\cite{GGV_WT}, hereinafter \wt, is a data structure for the representation of
sequences. Given a sequence $S[1,n]$ over an alphabet $\Sigma[1,\sigma]$, a \wt is
built as a perfectly balanced binary tree that subdivides the elements in the sequence
according to the position of their symbol in $\Sigma$. At the root node of
the tree, the original alphabet is divided into two halves $\Sigma_L$ and
$\Sigma_R$. Then, a bitmap $B[1,n]$ is associated with the node, so that $B[i] =
0 \iff S[i] \in \Sigma_L$, and $B[i] = 1$ otherwise. The process is repeated
recursively, considering in the left sub-tree the subsequence of symbols in
$\Sigma_L$ and in the right sub-tree the subsequence of symbols in $\Sigma_R$.
The tree has $\lceil\log \sigma\rceil$ levels, and a total of $n$ bits per
level, for a total size of $n \lceil\log \sigma\rceil$ bits.

The \wt supports several operations, being access and rank the ones we use in this work.
To access $S[i]$ in the \wt, we start at the root node of the tree and
check its bitmap $B$. If $B[i] = 0$, the symbol is in the left subtree. The
offset of the symbol in the left subtree can be easily computed as the number of
0s in the current node up to position $i$ (i.e. we compute the new $i$ as $
rank_0(B,i)$. Conversely, if $B[i] = 1$ we move to the right subtree and set $i
= i + rank_1(B,i)$. After $\lceil\log\sigma\rceil$ steps the path traversed
corresponds to the position of $[i]$ in the alphabet.
On the other hand, $rank_c(S,i)$ counts the occurrences of symbol $c$ in $S$ up to
position $i$ and it can be also implemented in $O(\log \sigma)$ time with a similar traversal.

In addition to plain representations of \wts, that essentially require
 the same space as the original uncompressed sequence, several other variants
 allow the compressed representation, achieving space proportional to the
 zero-order entropy of the input. In particular, we use a run-length compressed representation of the
 bitmaps of the \wt~\cite{MakinenN05} to exploit the regularities of our sequences.

\Section{Our proposal}

\SubSection{Formulation of the problem}
\vspace{-1mm}

Consider a sequence $S$ of data where each datum is a multidimensional entry,
from a given set of dimensions, of the form $(d_1, d_2, \dots, d_k) \in D_1
\times D_2 \times \dots \times D_k$.
This sequence may be a set of measurements, entries in a database, or any other
source of multi-dimensional data. We also may assume that the order of the
elements in the sequence could be random or according to some ordering,
either time or any other value related to one or more dimensions $D_i$.

The queries of interest in this domain involve aggregations of values for
specific values, or ranges of values, in one or more of the dimensions of the
dataset. That is, we are interested in counting the number of entries in the
sequence for a given value of $D_i$, or for a given set of constraints
across multiple dimensions, such as $d_1 = x$, $d_2 \in [d_{2\ell}, d_{2r}]$,
etc.

A plain representation of the original sequence is quite inefficient to retrieve
any kind of aggregate information. In order to answer these queries, additional
data structures are used to keep track of the specific values, thus
storing only accumulated data. Hence, both the representation of the sequence
and the additional data structures are required to keep the ability to access
and decode the sequence of operations.

\SubSection{Wavelet tree composition for aggregate queries}
\vspace{-1mm}
Our proposal is based on using \wts to represent the sequence and
provide efficient aggregation on a subset of the dimensions. The \wt
representation of a sequence provides a reordering of the elements in the original sequence according to the ``alphabet
order''. In other words, when a \wt of the sequence is built on an
alphabet $\Sigma$, the elements of the sequence are stably-sorted by their
position in this alphabet, so that the sequence can be accessed in the original order, using the root of the \wt, or in alphabet order, on the leaves.

Consider a sequence of tuples $(d_1,\dots,d_k)$ sorted by
$D_1$, then $D_2$, and so on. If we build a \wt, $S_{D_i}$, on this sequence
according to $D_i$, we are in practice stably-sorting all the elements of the
sequence by $D_i$. Hence, this \wt provides the support to efficiently
restrict our queries to specific values of $D_i$, thus effectively providing
aggregation capabilities in dimension $D_i$. For example, in order to count the number of
entries for a given $c \in d_i$, we just need to compute $rank_{c}(S_{D_i},n)$.

In order to provide aggregation capabilities in another dimension, we
can easily repeat the process, creating a \wt over the re-ordered
sequence using a different dimension.
If we build a new \wt, $S_{D_j}$, using the values of $D_{j}$, we are in practice grouping
elements according to that dimension, while keeping the previous ordering by
$D_i$. Using rank operations, we can now easily compute $rank_{c'}(S_{D_j},n)$ to
count the occurrences of any $c' \in D_{j}$. In addition, we can easily restrict the
previous query to the range covered by any $c$ in $D_i$, thus we can also
answer queries involving any pair of values in both dimensions.

Notice that, since after reordering by a second dimension the data are still
sorted by the previous one in each leaf (that is, they are sorted by $D_{j}$,
and by $D_i$ inside each group of $D_{j}$), we can also efficiently look for
the elements in the sequence that have a specific value for $D_{j}$ and a
range of values of $D_i$: we just need to restrict the search in the second
\wt to the range of values for the other dimension, which can be easily
computed. Table~\ref{tbl:reorderings} shows an example in which the data are originally sorted by day,
then reordered by activity and finally by employee.

\begin{table}[!h]
\begin{center}
\begin{scriptsize}
\begin{tabular}{ccc|ccc|ccc}
\hline
\multicolumn{3}{c|}{\textbf{Original order}}&\multicolumn{3}{c|}{\textbf{Reordering by Act}}&\multicolumn{3}{c}{\textbf{Reordering by Emp}}\\
\textbf{Day}&\textbf{Emp}&\textbf{Act}&\textbf{Day}&\textbf{Emp}&\textbf{Act}&\textbf{Day}&\textbf{Emp}&\textbf{Act}\\
\hline
1&1&C&1&1&A&1&1&A\\
1&1&E&2&1&A&2&1&A\\
1&1&A&1&2&B&1&1&C\\
1&2&E&1&1&C&2&1&C\\
1&2&B&2&1&C&1&1&E\\
2&1&A&1&1&E&1&2&B\\
2&1&C&1&2&E&1&2&E\\
2&2&E&2&2&E&2&2&E\\
3&2&E&3&2&E&3&2&E\\
\hline
\end{tabular}
\end{scriptsize}
\end{center}
\vspace{-1cm}
\caption{Example of reorderings with dimensions \underline{Day}, \underline{Emp}loyee and \underline{Act}ivity}
\label{tbl:reorderings}
\end{table}
\vspace{-0.5cm}

Our representation can efficiently compute queries involving consecutive elements in any of the reorderings. For example, count the number of entries involving activity E, which are 4, can be answered in the first reordering. In the same reordering, the query can be also restricted to a range of days (e.g. entries involving activity E during days $[1..2]$, which is 2). As for the second reordering, we may count the number of entries in which employee 2 was performing activity E during days $[1..2]$, which is 2.

Interesting enough, the different \wts need not be just stacked,
but can be combined in different ways according to the relevant queries in the domain. For
instance, if we have a 4-dimensional  dataset, over $A \times B \times C \times
D$, we may first process the sequence of values of $D$, sorting it by that
dimension. The result can then be used in two different \wts, one that
uses $C$ as the vocabulary and another that uses $B$. In this way, we shall be
able to easily count occurrences for a given $d$, a given pair $(c,d\star)$ or a
given pair $(b,d\star)$, where $x\star$ denotes the ability to search for a
specific value or a range of values.
An alternative representation that builds the 3 \wts in sequence, would be able to answer aggregation queries for given $d$,
$(c,d\star)$ or $(b,c,d\star)$. In this way, the final representation can
include as many combinations as needed to provide the necessary query
capabilities, at a cost of extra space. Given a set of queries, the problem of computing the minimum number of necessary \wts is an extension of
the shortest common superstring problem~\cite{RaihaU81}, i.e. obtaining the shortest string that contains a set of given strings.

\SubSection{Compressible sequences}


In many real-world scenarios, sequences of events are highly repetitive in terms of number of repeated $k$-tuples. This effect, called \textit{locality}, is increased with the granularity of the sequence. In the example described in the Introduction, if activities are recorded every minute, the sequence will have more locality than if they are recorded every hour. We propose now two implementations that take advantage of this property, providing solutions that are insensitive to the granularity of the sequences. This is interesting because it makes them suitable to represent sequences with a high level of detail, i.e. resolution, in little space and supporting aggregation queries.


We assume that the original sequence is sorted in such a way that aforementioned locality exists (e.g. time or space depending on the domain). Then, locality produces runs of symbols in the sequence. An out-of-the-box solution to exploit the existence of these runs is to use run-length compressed \wts~\cite{MakinenN05}. We refer to this solution as \wtrle. Runs on the sequence produce runs on the bitmaps of the \wt, which can be represented in few space and still support rank/select operations~\cite{DelprattRR06}. Overall, these \wts require space proportional to the number of runs in the sequence, instead of to the length of the sequence itself. Counting and access queries can be implemented as rank and access operations on the \wt, respectively. When queries involve more dimensions, an additional component is necessary. In the \wt composition explained above, a query on a \wt is refined on a second \wt, and so on. To do that, we need to store for each leaf of the \wt the number of elements that are lower, in the order defined by such \wt, than its corresponding symbol. This component is used to restrict the query in the subsequent \wt and it can be implemented as a plain array or as a sparse bitmap, depending on the size of the dimension.

\begin{figure}[]
\includegraphics[]{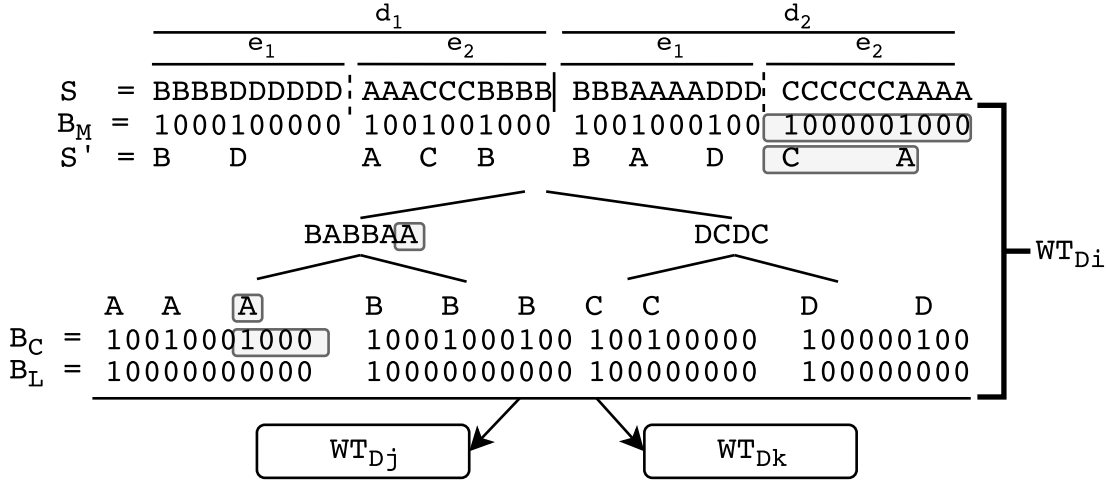}
\caption{Schema of the \wtmap solution for a sequence of 2 days, 2 employees, 4 activities and 10 time instants per day. Dimension $D_i$ represents activities. Highlighted elements are those visited to count the time-instants devoted by $e_2$ to activity $A$ during $d_2$, which are 4.}\label{fig_wt}
\end{figure}

This simple approach represents the runs at each level of the \wt, which is somehow a waste of space. Thus, our second approach, \wtmap, uses a technique to \textit{remove} the runs from the original sequence. We store a bitmap $B_M$, of length $n$, that marks the start of each run. Then, a \wt is built over a sequence $S'$ of length $n' < n$, in which each run of value $v$ is represented as a single value $v$. As this sequence does not contain runs, the \wt is built using plain bitmaps, which have the additional benefit of being faster for querying. A query on the original sequence can be easily mapped to a query on this new \wt using rank operations on $B_M$. However, this \wt does not contain information about repetitions of symbols, which is necessary to support counting operations. Hence, a second bitmap $B_C$ is built, storing in unary the length of each run. This bitmap considers the runs in the reordering performed by the \wt, i.e. it is aligned with the leaves of the \wt. Rank/select support is also necessary on this bitmap to solve counting queries. The two bitmaps, $B_M$ and $B_C$, are sparse, which is directly related with the existence of runs on the original sequence, and are implemented using Elias-Fano representation~\cite{OkanoharaS07}. The same final component described for the \wtrle is necessary here to do the composition between \wts. This component is shown as bitmap $B_L$ in Figure~\ref{fig_wt}, which illustrates this approach.

\Section{Experimental evaluation}

We design a set of experiments focused on a realistic use case, involving the storage and access to employee tracking information. We assume that we have a set of entries storing the following information: day, employee, time instant and activity. We want to keep track of all the information of our employees, hence being able to recover the specific activity that was being performed at a given time instant by a specific employee. In addition, typical queries in this domain involve checking the amount of time devoted to a specific activity by an employee, or by all employees. Hence we need to store the original sequence but also to efficiently answer aggregate queries. We shall consider for instance aggregations of time per activity/employee/day, and per activity/day.

\SubSection{Problem setup}

In this kind of data, time is a natural ordering of entries, and also provides compressibility due to locality: consecutive entries for the same employee and day should usually return the same activity, especially if the measurement frequency (resolution) is high. In this sense, storing the sequence of activities sorted by employee and time should lead to long runs of similar activities. We will consider the sequence of activities as sorted first by day, and then by employee and time, since days are a natural way of grouping data, the locality of activities is not affected and it becomes easier to answer queries for a specific day. Even though not all employees work every day, we can ignore this problem by storing every possible combination of day, employee and time instant, and using a special activity 0 to denote the fact that the employee was not working at that point. This increases the length of the sequence, but it makes computations simpler, and our representations will compress these long runs of values in little space. 

The queries that we will consider in our experiments involve aggregation operations and retrieval of specific entries in the sequence. We name our counting queries C-\textit{x}D-\textit{y}E-\textit{z}A, where D, E and A are days, employees and activities, and \textit{x,y,z} can denote a single value (1), a range (``r'') or all possible values (``a''). Hence, C-1D-1E-1A counts the time devoted on 1 day, by 1 employee, to a specific activity. A more complex aggregation is C-1D-aE-1A (1 day, all employees, 1 activity), that can be trivially extended to a range of days. We also test ranges of days with query C-rD-1E-1A. In addition to these aggregation queries, we also test the ability of our proposal to \textit{access} specific positions of the original sequence: query \textit{Acc} retrieves the activity being performed, given an employee, day and time instant. 

Taking into account the characteristics of the dataset, we build our representation sorting entries by activity first, in a first \wt decomposition; then we group again by employee. The first decomposition allows efficient counting operations by activity and day (or range of days); the second one works for queries on activity-employee-day. Notice also that since we actually store every possible combination of values, an \textit{Acc} query can be easily translated into a position to extract in the sequence.

\SubSection{Baseline representation}

As a baseline for comparison we build a simpler representation that uses the same ordering of the information and takes advantage of it to answer the same relevant queries with additional data structures. The baseline consists of two separate components: a representation of the original sequence, which can be used to access the original data, and an additional data structure for aggregation queries. This baseline thus follows the usual approach of separating the original data and precomputing aggregated values in a separate structure.

The sequence of tuples is stored using two arrays and two additional bitmaps. Initially, the sequence is ``compressed" removing repeated entries: each group of consecutive entries with the same values is replaced with a single entry storing the activity and the
amount of time devoted to it. The compact representation works with this shorter sequence. 
A bitmap $B_D$ marks the positions in the sequence where a change of day occurs; a second bitmap $B_E$ is used to locate positions where each employee sequence begins. Using these two bitmaps, we only need to store activities and times for each position: two arrays store the activities and time values for each entry. With this representation the original contents can be recovered using the bitmaps to locate the appropriate run and processing the sequence of times to locate the desired position. Operations that require counting (time spent in an activity by an employee, or by all employees in a day) can also be computed using this sequence, but may require the traversal of several different regions of the array to compute the overall result.

To provide efficient counting operations, the baseline also stores a data structure specifically designed to compute aggregated values. The data structure we use is called CMHD~\cite{Brisaboa2016}, and is a compact representation designed for aggregation queries in hierarchical domains such as OLAP databases. This representation essentially decomposes the $n$-dimensional datasets, according to predefined hierarchies, and is able to store accumulated values at each level of decomposition (for instance, given products and places, it can store in the same representation accumulated values by category and country, and also by individual products and places). In our domain no natural hierarchies appear apart from  days and time. Nevertheless, we can build fictitious hierarchy levels to efficiently answer queries: decomposing first some dimensions we can  obtain a first level in the CMHD that stores cumulative values for all employees, 1 activity and 1 day; then, decomposition can continue in a second level to store cumulative values for 1 employee, 1 activity, 1 day. In this way, queries that involve a single value or all values can  be answered efficiently, as long as a hierarchy is appropriately built in the CMHD.

\SubSection{Experiments and results}

We built several synthetic datasets following the expected distribution for the domain. All our datasets have similar realistic characteristics but different size in different dimensions, to measure the evolution of the proposals. The changing parameters in the datasets are the number of activities \textit{A}, the number of employees \textit{E} and the number of time instants per day, or resolution, \textit{R}. The number of days is set to a fixed value (500), since it has little effect on query times and only changes the length of the sequence. Datasets \textit{T-rrr} (where \textit{rrr} is the resolution) are built with $E=50$, $A = 16$ and $R \in \{720,1120, 5760, 11520\}$. Datasets \textit{Dat-ee-aa} are built with $R=5760$, $E \in \{20,50,100\}$ and $A \in \{16,32,64,256\}$. In all the datasets, we consider that employees work on shifts (50\% of the time each working day), and have free days (only 80\% of the employees work each day). For each dataset and query type we generated sets of 1 million queries, selecting random values and intervals. 

In this section we compare our two proposals, \wtrle and \wtmap, with the alternative  \baseline representation, for all the test datasets. All the representations are implemented in C++ and compiled with g++ with full optimization enabled. All the experiments in this section were performed in an Intel Core i7-3770@3.60GHz and 16GB of RAM, running Ubuntu 16.04.4.

Due to space constraints, we only show space required by some of the datasets,
for selected numbers of employees, activities and resolutions, since behavior in
all the datasets is similar. Table~\ref{tbl:espacio} shows this summary. For
each dataset, the input size is computed considering a completely plain
representation of all the tuples in the sequence, with each value stored using
32 bits, so this size increases linearly with the number of entries. All the compressed
representations are much smaller than the original sequence, thanks to the efficient compression of the runs of repeated values. Both of our representations are in all cases considerably smaller (5 to 10 times smaller) than the baseline. The most efficient representation is \wtmap, since it removes runs of repeated values in a single step, while \wtrle has a small overhead to compress the runs in every level of the \wt.

\begin{table}[!h]
\begin{center}
\begin{scriptsize}
\begin{tabular}{crrrr}
\hline
Dataset 	& Input size	& \baseline	&	\wtrle	& \wtmap	\\
\hline
T-2880 &  1,098.63  &  30.73  &  3.58  &  \textbf{3.33}\\
T-11520 &  4,394.53  &  30.85  &  4.09  &  \textbf{3.70}\\
\hline
Dat-20-16 &  878.91  &  12.41  &  1.50  &  \textbf{1.35}\\
Dat-20-64 &  878.91  &  13.20  &  2.04  &  \textbf{1.51}\\
Dat-20-256 &  878.91  &  13.74  &  2.35  &  \textbf{1.65}\\
\hline
Dat-50-16 &  2,197.27  &  30.65  &  3.82  &  \textbf{3.51}\\
Dat-50-64 &  2,197.27  &  33.14  &  5.35  &  \textbf{3.99}\\
Dat-50-256 &  2,197.27  &  34.43  &  6.20  &  \textbf{4.35}\\
\hline
Dat-100-16 &  4,394.53  &  61.76  &  7.72  &  \textbf{7.27}\\
Dat-100-64 &  4,394.53  &  65.78  &  10.78  &  \textbf{8.15}\\
Dat-100-256 &  4,394.53  &  67.96  &  12.45  &  \textbf{8.84}\\
\hline 
\end{tabular}
\end{scriptsize}
\end{center}
\vspace{-1cm}
\caption{Space required by all the datasets (sizes in MB)}
\label{tbl:espacio}
\end{table}
\vspace{-3mm}

Next we compare our proposals with the baseline for the different queries
described for this domain. Again, due to space constraints, we only show the
results for some of the studied datasets. Comparison results for the other
datasets used are similar to those introduced here, with query times varying
slightly depending on the size of the sequence and the different dimensions.

Table~\ref{tbl:times} contains the query times for all approaches for all queries.
The first query we study is the \textit{access} query Acc, that recovers random positions in the sequence. In
this query both of our proposals are faster than the baseline in all cases, and
again \wtmap is the fastest of our alternatives.

\begin{table}[!h]
\begin{center}
\begin{scriptsize}
\setlength{\tabcolsep}{3pt}
\bgroup
\begin{tabular}{c||ccc||ccc|ccc|ccc}
Dataset 	& \multicolumn{3}{c||}{Acc} & \multicolumn{3}{c|}{C-1D-1E-1A} & \multicolumn{3}{c|}{C-1D-aE-1A} & \multicolumn{3}{c}{C-rD-1E-1A} \\ 
			& base &	\wtrle & \wtmap	& base &	\wtrle & \wtmap	& base &	\wtrle & \wtmap	& base &	\wtrle & \wtmap	\\
\hline
Dat-20-16 	& 0.51	& 0.36	& \textbf{0.23}	& 1.46 (\textbf{0.45}$^\ast$)	& 1.12	& 0.74	& \textbf{1.35}	& 2.21	& 1.67	& ---	& 7.25	&	\textbf{3.49} \\
Dat-50-64 	& 0.56	& 0.53	& \textbf{0.32}	& 1.57 (\textbf{0.58}$^\ast$)	& 1.58	& 1.15	& \textbf{1.38}	& 3.30	& 2.28	& ---	& 8.60	&	\textbf{4.65} \\
Dat-100-256	& 0.61	& 0.56	& \textbf{0.46}	& 1.50 (\textbf{0.66}$^\ast$)	& 1.97	& 1.57	& \textbf{1.31}	& 4.50	& 2.97	& ---	& 11.78	&   \textbf{7.19} \\  
\multicolumn{13}{l}{$^\ast$ Values in parentheses are performed counting sequentially in the baseline sequence representation}
\end{tabular}
\egroup
\end{scriptsize}
\end{center}
\vspace{-1cm}
\caption{Query times for access (Acc) and counting queries. Times in $\mu$s/query}
\label{tbl:times}
\end{table}
\vspace{-3mm}

Next, also in Table~\ref{tbl:times}, we study aggregation queries. For C-1D-1E-1A we show two different times for the baseline: the first time is obtained querying the CMHD representation, and the time in parentheses is obtained traversing the sequence representation. In this query, the naive approach that traverses the sequence obtains in practice the best results. This is due to the relatively small number of activities per day in this domain. Note however that the cost of the naive sequence traversal is linear on the number of entries to search, so it may be slower than any of the indexed proposals even for simple queries if the number of entries to search is relatively large. In any case, our representations are still competitive in query times and use up to 10 times less space. Moreover, our proposals are faster than the indexed CMHD representation in many cases.

In the last two queries a naive traversal of the sequence in the baseline becomes too slow (5 to 20 times slower than our proposals) so we only compare with the CMHD. In query C-1D-aE-1A the CMHD can be faster by storing cumulative values for the corresponding query, but our proposals are still competitive. In addition, our proposals can answer a generalization of the query to a range of days (C-rD-aE-1A) in roughly the same time, whereas the CMHD would have to resort to summing up multiple entries, becoming much slower. This is also shown in query C-rD-1E-1A: our proposals are a bit slower in this query, but query times are not too high in comparison with previous queries; however, the CMHD would have to obtain and sum cumulative values for each day, becoming too slow in practice, with times again comparable to a naive traversal of the sequence. A partial improvement on the CMHD could be obtained by generating fixed divisions as time hierarchies (by week, by month), reducing the number of sums to perform, but this requires previous knowledge of the desired ranges, and unless queries involve ranges exactly matched with hierarchical values, the CMHD is expected to be much slower in these queries. While the CMHD is much more dependent on a previous definition of hierarchies of interest,  our proposals are more flexible and can efficiently answer range queries on the different dimensions.  

\Section{Conclusions}

We have presented two different compact representations for multidimensional sequences with support for aggregation queries. We show that our representations can take advantage of locality to store very large datasets in reduced space. Our experiments show that our proposals are smaller, and faster to access, than simpler representations of multidimensional sequences. Also, and even if storing much more information, our proposals are competitive on aggregation queries with state-of-the-art data structures designed specifically for those queries.   

An interesting line of future work is to explore the effect of the encoding of the symbols in real domains. In our running example, some activities may be more likely to be performed after a specific one than others. For example, take-a-nap activity after lunch. By encoding \textit{similar} activities with closer symbols, these regularities are automatically exploded by the \wtrle, but not for any other of the solutions.

\Section{References}
\bibliographystyle{IEEEtran}
\bibliography{biblio}

\end{document}